%% file: article.tex
\documentclass[sigconf]{acmart}

\usepackage{booktabs} 

\setcopyright{acmcopyright}

\acmDOI{xx.xxx/xxx_x}

\acmISBN{978-1-4503-9517-5/23/03}

\acmConference[SAC'23]{ACM SAC Conference}{March 27 –April 2, 2023}{Tallinn, Estonia}
\acmYear{2023}
\copyrightyear{2023}

\acmArticle{4}
\acmPrice{15.00}

\begin{document}
\title{Control-Flow Integrity at RISC:\\ Attacking  RISC-V by Jump-Oriented Programming}
  
\renewcommand{\shorttitle}{Control-Flow Integrity at RISC: Attacking  RISC-V by JOP}

\author{Olivier Gilles}
\orcid{0000-0002-3776-2071}
\affiliation{%
  \institution{Thales Research \& Technology}
  \streetaddress{adresse}
  \city{Palaiseau} 
  \country{France} 
  \postcode{91002}
}
\email{olivier.gilles@thalesgroup.com}

\author{Franck Viguier}
\affiliation{%
  \institution{Paris-Sorbonne University}
  \streetaddress{adresse}
  \city{Paris} 
  \country{France} 
  \postcode{75006}
}
\email{Viguier.Franck@outlook.fr}

\author{Nikolai Kosmatov}
\orcid{0000-0003-1557-2813}
\affiliation{%
  \institution{Thales Research \& Technology}
  \streetaddress{adresse}
  \city{Palaiseau} 
  \country{France} 
  \postcode{91002}
}
\email{nikolai.kosmatov@thalesgroup.com}

\author{Daniel Gracia P\'erez}
\orcid{xxxx-xxxx-xxxx}
\affiliation{%
  \institution{Thales Research \& Technology}
  \streetaddress{adresse}
  \city{Palaiseau} 
  \country{France} 
  \postcode{91002}
}
\email{daniel.gracia-perez@thalesgroup.com}

\renewcommand{\shortauthors}{Gilles et al.}

\begin{abstract}
RISC-V is an open instruction set architecture recently developed for 
embedded real-time systems. To achieve a lasting security on these
systems and design efficient 
countermeasures, a better understanding of vulnerabilities to novel 
and potential future attacks is mandatory. This paper demonstrates that  
RISC-V is sensible to Jump-Oriented Programming, a class of complex 
code-reuse attacks, able to bypass existing protections. 
We provide a first analysis of
RISC-V systems' attack surface exploitable by such attacks, and show how 
they can be chained together in order to build a full-fledged attack.
We use a conservative hypothesis on exploited 
registers and instruction patterns, in an approach we called 
\emph{reserved registers}.
This approach is implemented on a
vulnerable 
RISC-V application, and successfully applied to expose an AES256 secret. 

\end{abstract}

%
%


\maketitle

\input{intro}
\input{background-cra}
\input{attack_surface}
\input{craft_jop}
\input{experiment}
\input{related}
\input{conclusion}

\bibliographystyle{ACM-Reference-Format}
\bibliography{bibliography} 

\end{document}

%% file: intro.tex
\section{Introduction}\label{sec:intro}

RISC-V\footnote{https://riscv.org} is an open Reduced Instruction Set Computer (RISC) architecture mostly targeting embedded and real-time
systems. While reduced instruction set architectures innately have a smaller
attack surface than Complex Instruction Set Computer (CISC) architectures, many of them run critical
systems, including Industrial Control Systems (ICS) or Cyber-Physical Systems
(CPS), whose failure may have dramatic consequences for the physical environment,
including environmental disasters and loss of human lives.

Using a novel Instruction Set Archiecture (ISA) brings several benefits. 
Beyond security advantages by
taking into account the experience on the latest attacks, 
one major interest is its open status, where trust in the architecture relies on 
community review. This also enables national independence in microchip supplies,
a very important feature as target systems may be strategical, and export
restrictions become more common.

While most RISC-V architectures  offer a satisfying level of security
compared to similar classes of systems~\cite{Lu21,Nicholas20}, they
will increasingly become the target to complex attacks as their relevance in the
industrial and strategical field increases. Eventually, state-backed attackers
are bound to try and attack them. In order to anticipate this threat, security
researchers face the challenge to anticipate potential vulnerabilities 
and imagine suitable protection mechanisms.

Code-Reuse Attacks (CRA) and specifically Jump-Oriented Programming are amongst the
most complex attacks to realize, but also to prevent. They can be very
powerful when successful, as they can allow the attacker to run an arbitrary
sequence of instructions within the corrupted application. In this document we
adopt the attacker's point of view and try to perform a Jump-Oriented Programming (JOP) attack, with
the intent of (1) better understanding the vulnerabilities of RISC-V systems, and
(2) ultimately designing better countermeasures to prevent these attacks.

\emph{Contributions.} We summarize our contributions as follows:
\begin{itemize}
\item a first analysis of vulnerabilities to JOP attacks on RISC-V architecture;
\item a description of how said vulnerabilities may be exploited in a JOP chain;
\item a demonstration of feasibility by implementing and 
testing a JOP attack on a vulnerable RISC-V application.
\end{itemize}

\emph{Outline.}
Section 2 briefly presents existing work on Code-Reuse Attacks. Section 3
describes vulnerabilities with RISC-V instructions, 
extensions and Application Binary Interface (ABI).
Section 4 explains how these vulnerabilities can be exploited to craft a
JOP chain, while Section 5 describes an attack that
we developed in order to corrupt a vulnerable RISC-V application. 
Section 6 compares our approach to other efforts related to RISC-V security, regarding 
either attacks or defenses.
Finally, Section 7 provides a conclusion.

%% file: background-cra.tex
\section{Background on Code-Reuse Attacks}\label{background}
\label{introcra}

The aim of a Code-Reuse Attack (CRA) is to take control of an application
execution through the use of existing functions and system calls within the
target application, in order to perform unintended actions (typically, sabotage
on Cyber-Physical Systems) or to leak secret information.
Indeed, CRA may also be used as part of a more complex attack involving
different exploits.

CRA is not in itself a primary attack, but relies on an earlier memory
corruption, typically on the stack or the heap. Such attacks are well-known, but still
prevalent in many systems~\cite{Younan04codeinjection}. The originality of CRA
comparatively to regular code injection (e.g. shellcode injection) is that,
instead of redirecting the execution flow toward injected code (generally in a buffer),
it redirects the execution  toward existing code in the application
in order to obtain a malicious effect. A simple example of such attacks is
return-to-libC~\cite{Solardesigner97}, where the execution flow is redirected to
a single function after manipulation of arguments within the stack of the
corrupted function. The redirection is made by overwriting the return address,
contained in the stack. More sophisticated attacks with the same principle of stack
corruption have emerged, among which the most notorious being what is
commonly called Return-Oriented Programming (ROP)~\cite{Shacham07,Carlini14}.
It consists in using \emph{gadgets chains}, an assembly of code snippets found
within the target application, and which finish by a linking instruction. In
the case of ROP, the linker instruction will be a return to caller instruction, which will
pop the corrupted stack and then redirect the flow to the next gadget, and so on.
Using this approach, the attacker can run an arbitrary sequence of legit
instructions, effectively running a malicious, potentially a Turing-complete~\cite{Tran11}
application within the application.

\subsection{Countermeasures}

Multiple methods were proposed and used in order to defend against return-to-libc
and JOP. Address Space Layout Randomization (ASLR) makes the attack more
difficult to craft, as the attacker needs to guess the address of target function
or gadgets. However its efficiency has been proven limited, especially in 32-bits
architectures~\cite{Shacham05}. Stackguard~\cite{Cowan98} introduced the notion
of canaries to protect the integrity of the management data (including the return
address) in the stack, yet solutions relying on secrets depend much on the system
entropy, and tend to decline as the system uptime increase -- an important issue
in embedded systems that can run for decades without reboot.

Abadi et al.~\cite{abadi05}
propose a solution named Control-Flow Integrity (CFI), which consists in (1) analyzing the
Control Function Graph (CFG) from the protected application and (2) check at runtime
each jump and return instruction follows legit CFG edges. CFI makes
theoretically all kinds of CRA nearly impossible to implement, and does indeed
stops most ROP attacks if implemented with a shadow stack~\cite{carlini15}, although
often leading to significant fall of performances~\cite{Burow17}. Its
efficiency, however, is limited against attacks on the forward edges of the CFG
(as opposed to ROP which hijacks backward edges). The precision of the CFG also limits
its efficiency, a weakness that was exploited to overcome this defense~\cite{carlini15,goktas14}.
This led to new classes of attacks: Call Oriented Programming (COP) and more generally,
Jump Oriented Programming~\cite{bletsch11} (JOP). Specific solutions addressing return-to-libc and ROP
attacks for RISC-V architectures have been proposed by using pointer authentication
at hardware level~\cite{Wang22_rettag}, or by using an encrypted shadow stack in~\cite{bruner21}.

\subsection{Overcoming CFI: Jump-Oriented Programming}

\begin{figure}
\includegraphics[scale=0.4]{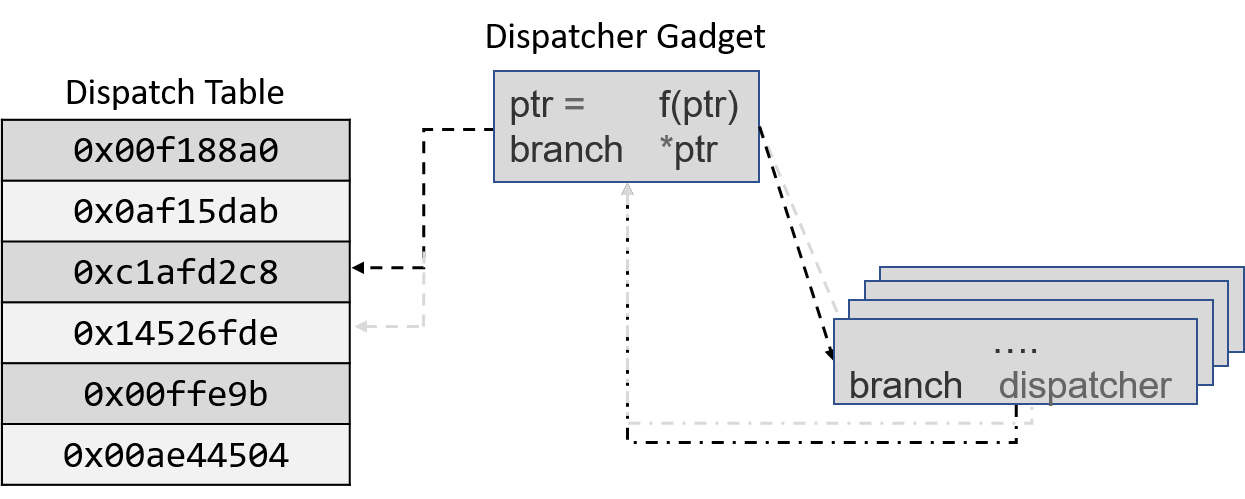}
  \caption{Dispatcher gadget role in JOP}\label{fig:jop_dispatcher}
\end{figure}

Much like ROP, JOP consists in assembling code gadgets containing
useful instructions in order to execute a malicious sequence of instructions
present within the target application. The linking instruction 
allowing to chain these gadgets,
however, is not the return instruction but any instruction performing an indirect
jump (i.e. a jump on an address provided by a register). While in the case of ROP
attacks the target gadgets' addresses are stored in the stack, and the poping of these value
is handled by the function epilogue, in the case of JOP it must be done by a
specific gadget, named \emph{dispatcher gadget}. This gadget will load the addresses
of \emph{functional gadgets} with a table in memory, generally injected into a
buffer, and then jumps to said address. Each functional gadget must end with a jump
to the gadget dispatcher, as illustrated in Figure~\ref{fig:jop_dispatcher}. The
most prominent example of JOP publicly known to this day is a complex rootkit targeting
x86/Linux2.6, bypassing most recent defenses~\cite{Cheng11}.

Assembling a JOP gadget chain is more complex than doing so for a ROP gadget chain,
as (1) the combination in the choice of registers and (2) the side-effects in the gadget
may break the gadget chain management, for instance by overwriting the registers
containing the addresses of the dispatcher gadget or the gadget table. Since
gadgets research and chaining is highly dependent on both ISA and ABI, feasibility
in new architectures is not proven. In the next sections, we demonstrate the
feasibility of JOP attacks on applications compiled for the RISC-V architecture.

%% file: attack_surface.tex
\section{RISC-V attack surface}\label{sec:attack-surface}

JOB attacks rely on the ISA and ABI of the target, and as such they
represent the attack surface of the target.
In this section we present the RISC-V ISA and ABI elements and properties of
interest for the realisation of JOB attacks.

Note that the ABI is purely a convention, i.e. while an application on a RISC-V
processor must follow the RISC-V ISA specificiations, it is not mandatory to
follow any particular ABI.
In this article, we made the hypothesis that the target application binary and
all related software artifacts were generated (e.g. compiled) following the
convention described in Chapter 18 of the RISC-V specification\cite{rvspecs}.

\subsection{ISA: Instruction Set Definition}
\label{ref:riscv-isa}

The RISC-V ISA defines multiple extensions to the base instruction set.
Amongst these extensions, only the Base Integer ISA (RV32I) is mandatory in all
RISC-V implementations. Depending on supported extensions, new registers
will be available (e.g. floating registers). A 16-bits compressed version
of Base Integer ISA is also available. Available registers in the target
architecture will strongly impact its attack surface relatively to JOP
attacks, as more gadgets will be eligible when more registers are present.

In our experiment, we had the most possible conservative approach by only
considering the Base RISC-V ISA and the Base Integer ISA, thus exploiting a 
minimal attack surface. Any added extension will increase the likelihood of 
finding gadgets within the target application.

\subsection{ISA: Instructions Alignment}

The number of gadgets available is increased if the instructions are not
aligned, since in the latter case, instructions may be loaded after their
intended beginning, and thus interpreted with unintended opcodes and
arguments~\cite{Jaloyan20}. Contrary to CISC architectures such as
x86, RISC-V being a RISC architecture, RISC-V instructions have a fixed size.
However, an extension providing compressed instructions (RV32C and RV64C)
is defined in the standard, although not mandatory. RISC-V implementations
supporting this extension may be target to instruction offset shifting
attacks, where the attacker jumps after the begining of an instruction and
thus execute unintended instructions.

\subsection{ISA: Instructions of Interest}

\begin{figure}
\includegraphics[scale=0.6]{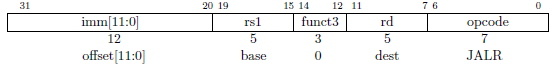}
\caption{RISC-V Jump-And-Link-Register instruction}\label{fig:cva6jalr}
\end{figure}

Designing a JOP attack relies on chaining functional gadgets. This chaining is
performed by using control transfer instructions exploiting registers as jump
target designator. In the RISC-V Base Integer ISA, there is only one instruction
allowing such operation: the Jump-And-Link-Register instruction (JALR), as 
illustrated in Figure~\ref{fig:cva6jalr}.

JALR computes the target address for the control transfer by adding a 12-bits
signed immediate to the rs1 register, then setting the least-significant bit
of the result to zero, allowing to reach +/- 2 ko from the source address.
Reachable addresses can be further extended by using the Add Upper Immediate 
to PC (AUIPC) instruction, which add 20 bits to the memory, effectively making 
all the physical memory addressable with the immediate. In the case of a JOP 
attacks, exact target addresses are stored registers are stored into the 
dispatch table and loaded directly into \emph{rs1} (one register argument), so 
immediate is set to zero and thus no AUIPC instruction is needed by the attacker. 

\subsection{ABI: Registers of Interest}

JOP attacks exploit instructions allowing jumping to an address stored within a
register. The Base Integer ISA from RISC-V defines 32 generic registers usable in
user mode:
\begin{itemize}
\item a global pointer addressing different symbols (functions, global variables)
  within the application;
\item a return register containing the return
  address of the last caller function;
\item a stack pointer;
\item a thread pointer;
\item 8 argument registers, used to pass arguments to functions;
\item 6 temporary registers, used to store intermediary results;
\item 12 save registers, used by compiler to otimize access to non-volatile values  
  accessed from different functions.
\end{itemize}

Table~\ref{tab:rvabi} shows the registers mnemonics and usage purpose as defined
by the RISC-V specification ABI.
As shown in the table, the ABI states that return, argument and temporary
registers are the only registers saved by caller functions.

\begin{table}
\includegraphics[scale=0.75]{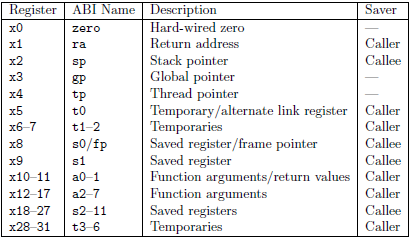}
  \caption{RISC-V registers usage as describe by ABI}\label{tab:rvabi}
\end{table}

\subsection{ABI: Convention on Function Calls}

Convention on RISC-V function calls favors passing arguments by registers
when possible. Eight integer registers are saved for this purpose in the RV32I. 
When no register is available, arguments are stored in the stack.

As mentioned in Section~\ref{introcra}, CRA in general and JOP in particular consist in
using existing functions and system calls within the target application.
Since the behaviour of these functions and system calls is controlled by their
parameters, those will be the favorite targets for an attacker. Two information
in particular are interesting regarding argument passing: (1) assignment of
registers and (2) layout of parameters' values within these registers. In the
next subsections, we use the XLEN term to describe the registers size in bits (32 or 64
according to the RISC-V architecture).

\subsection{ABI: Argument Registers Allocation}

Arguments with a size equal or less than XLEN are assigned to registers
according to their position within the function's signature, e.g. first
argument is assigned to first argument register (a0), second argument is
assigned to second argument register (a1) and so on. Arguments with a size
greater than 2*XLEN are passed by reference (so they occupy a single register).
Finally, arguments with a size greater than XLEN and less or equals to 2*XLEN use
a pair of registers. Hence in the case of manipulation of data belonging to the
latter group (such as \textit{long long} in a RV32I), more arguments
registers than actual arguments will be used. This makes the application more 
likely to use upper argument registers. Since the number of argument registers 
used in the application is a strong limitation in the attacker ability to find 
gadgets due to our approach (see~\ref{ref:craft_jop}), this pattern increases
its vulnerability against JOP.

\subsection{ABI: System Calls}
\label{ref:riscv-syscalls}

In RISC-V system calls are performed through the ecall instruction. Each
system function has an integer identifier allowing identification during the
call. This identifier must be set into the last argument register (a7) before
call execution. If any arguments are necessary to the system call, they must be
passed through the other argument registers (a0-a7).

While it is possible to manipulate a7 to change any syscall to the desired one, 
we found that their was not many occurrences of manipulation of a7 outside of 
setting of syscalls and without of adversary side-effects on lower argument 
registers. Hence, we rather identified within the application the system call 
of interest, which was easy in our case, since we target the commonly used 
\emph{write} syscall.

%% file: craft_jop.tex
\section{Crafting a JOP Chain for RISC-V applications}
\label{ref:craft_jop}

As mentioned in Section~\ref{ref:riscv-isa}, we adopted a conservative approach regarding
our selection of registers of interest. We used the same approach regarding their
exploitation: we defined a subset of register of interest to be \emph{reserved registers},
that are used by the attack to store addresses necessary to the continuity of the
attack. These registers are:
\begin{itemize}
  \item \emph{a3}, the gadget dispatcher address (which can consists in one or two
    addresses, depending on the kind if gadget dispatcher);
  \item \emph{a4}, the current entry in the dispatch table.
\end{itemize}

We did not consider in our experiment the possibility to save and restore these
registers' values, so any writing on the registers by any gadget after the gadget
initialization will interrupt the attack. To find a way to save/restore these
registers (either through memory or through direct register copy) would considerably
ease the finding of gadgets and the crafting of the JOP chain.

\subsection{JALR Instruction}

The Jump-And-Link-Register instruction may be generated by the compiler when the
code tries executing a function pointer. The specific case of JALR performed
on a parameter happens when said function pointer was passed to the parent
function through arguments -- in that case, JALR will jump to the address
contained in the related argument. Hence manipulating this argument allows
modifying the execution flow.

\begin{figure}
\includegraphics[scale=0.5]{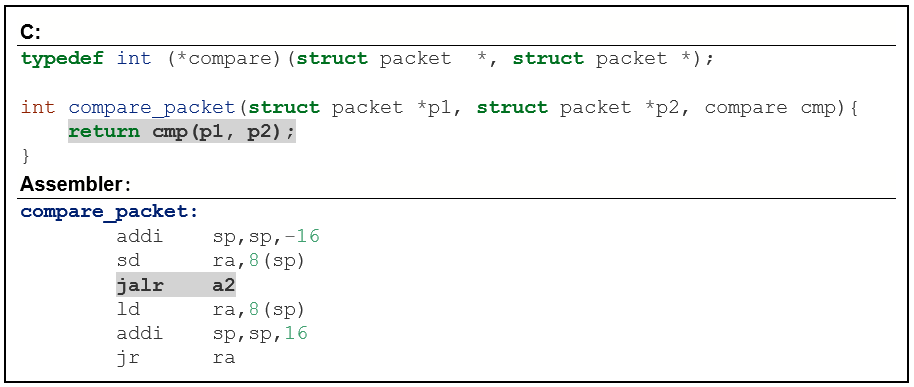}
\caption{Pattern generating vulnerable JALR}\label{fig:jalr}
\end{figure}

For example, a generic comparison function would take as arguments 2 structures
to be compared, and a pointer on a comparison function. The two structures will
occupy registers a0 and a1 (supposing they fit in), and the comparison function
address will be stored in register a2. When calling the comparison function, the
generic comparison function will use a JALR (depending on compiler options) and
branch to register a2, as illustrated in Figure~\ref{fig:jalr}. The generic function
is a good candidate for chaining JOP gadgets, as it allows taking control of
execution flow with the assumption that the arguments are already controlled by
the attacker.

\subsection{Dispatcher Gadget}\label{subsec:dispatcher-gadget}

The dispatcher gadget is an essential part of any JOP attack. Its behaviour is
similar to an instruction pointer, as it allows (1) loading an address in the
dispatch table, (2) jumping to said address and (3) moving to the next gadget
table entry. The dispatch table is typically injected in the application through
through the buffer that was used to initiate the attack. The execution
order of gadgets is generally sequential, but it may be more complex, according
to available dispatcher gadget candidates in the application. In any case, it is
defined by the dispatch table layout which is crafted by the attacker during the
weaponization phase of the attack (as defined in Lockeed Martin cyber kill 
chain\footnote{https://www.lockheedmartin.com/en-us/capabilities/cyber/cyber-kill-chain.html}).

\begin{figure}
\includegraphics[scale=0.7]{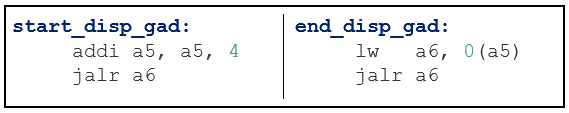}
\caption{2-stage gadget dispatcher}\label{fig:2stages}
\end{figure}

While its expected behaviour is simple, to find an ideal dispatcher gadget is a
difficult task. In the case of a RISC-V application, the attacker must find the
following pattern:
\begin{itemize}
  \item one instruction to increment a1 (supposing the dispatch table was
    stored there);
  \item one instruction loading a2 from a1 (were a2 will contain the next
    address to be jumped to);
  \item one instruction to perform an indirect jump (JALR) to the address stored
    in a2.
\end{itemize}
While other instructions may be interlacing with the instructions above, they must
not tamper with either a1 or a2 (neither any registers used in further
operations, such as the one containing the address of the gadget dispatcher
itself). While this issue may be solved by using other registers than argument
registers, another solution to address this issue is to use a 2-stage gadget
dispatcher, as proposed by~\cite{joprocket21}. In that case, the
first ``stage'' of the dispatcher gadget is effectively a gadget allowing to parse
the dispatch table (generally by incrementing a pointer), and then branches on the
second stage, which in turn performs a load on the dispatch table current position,
and then branches on this value using the JALR instruction. Figure~\ref{fig:2stages}
illustrates a 2-stage dispatcher gadget on RISC-V.

\subsection{Functional Gadgets}\label{subsec:func-gadgets}

The dispatcher gadget is only useful if it can address the functional gadgets
needed to perform the attack. There are different families of functional gadgets,
as described for ROP attacks by~\cite{bletsch11}:
\begin{itemize}
\item arithmetic and logic gadgets
\item memory access gadgets
\item function call gadgets
\item system call gadgets
\item branching gadgets
\end{itemize}
Even in ROP attacks, one or more gadgets of each of these families need
(e.g. memory access will require a loader and a saver gadget) to be present in
order to achieve Turing-completeness~\cite{Cheng11}.
In the case of JOP attacks were more than one register may be exploited to chain
the gadgets, the number of gadgets needed may increase even further.

Reaching Turing-completeness however, although intellectually satisfying, is
often not needed to perform an actual attack -- hence the question whether there
is enough functional gadgets available in an application to perform an attack is
hard to answer, particularly since the attacker may adapt the attack to available
gadgets. In our experiment we did not aimed for Turing-completeness, so we did not 
need all the gadgets types listed above. We needed, and were able to find (or 
rather to generate in our vulnerable application) (1) arithmetic gadgets, (2) 
memory access gagdets and (3) system call gadgets. 

In addition to these functional gagdets, we also needed a initializing gagdet, 
which is a one-shot functional gadget used to configure the registers containing 
the dispatcher gadget and the dispatch table addresses. While it is actually a 
memory access gadget in literature, its specific role makes it unique, as it will 
be the only gadget allowed to write in the reserved registers

\subsubsection{Arithmetic and logic gadgets}

\begin{figure}
\includegraphics[scale=0.7]{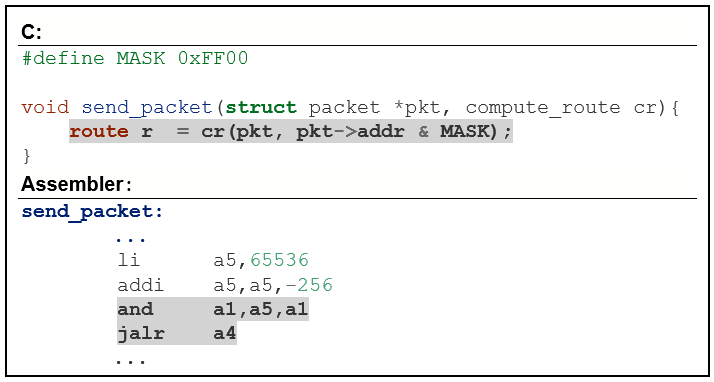}
\caption{AND gadget}\label{fig:andgadget}
\end{figure}

As suggested by their name, arithmetic and logic gadgets are used to perform
arithmetic and logic operations on registers' values. During our experiment, we
focused on argument registers, thus one need to find a gadget performing the
needed operations on the current function parameters. An example of AND operation
executed on arguments of a function is provided in Figure~\ref{fig:andgadget}.

\subsubsection{Memory access gadgets}\label{subsubsec:mem-access-regs}

Memory access gadgets are the most important gadget in any attack scenario -- second only to
system call gadgets. They allow reading and writing values from/into memory, and are always
necessary to parametrize said system calls, but also more generally to access
sensitive data. One simple example is to write in memory the name of a sensitive
file in file system (e.g. /etc/shadow in a Linux context) and then calling the
syscall \emph{open}, allowing tampering of said file (providing the target
application has root privilege) and thus allowing the attacker to modify users'
passwords. 

Memory access gadgets can be found, amongst other places, in functions manipulating
structures. As these structures' size is typically greater than XLEN*2, they are
passed by reference, and then a memory read will be performed. Conversely,
modification on the values of the structure parameters will be performed through memory
writings.

The location of writing and reading depends on the structure address, passed
through a register. Manipulating this register allows arbitrary reading or writing
within the memory (e.g. the dispatch table).

\subsubsection{System call gadgets}

As mentioned in Section~\ref{ref:riscv-syscalls}, system calls following the
RISC-V ABI use argument register a7 to pass the ecall identifier. Since argument
registers are assigned lower index first, finding functions manipulating a7 out of
system libraries is possible, yet very likely to lead to side effects on lower 
argument registers. In our research, we did not found gagdets manipulating a7 
without overwriting one of the reserved registers. Hence performing system calls 
practically means branching into related primitives in system libraries while 
ensuring a configuration of registers that will trigger the kernel action expected 
by the attacker.

Practically, that means we exploit entries into the Procedure Linkage Table (PLT).
This table and related functions are generated by the Linker during the linking
phase and points towards used functions in the libraries linked to the
application. From these addresses, we have an entrypoint to functions of interest
(e.g. open in libc), and then we can jump up to a fixed offset of the function 
with the instruction of interest (in our case, the instruction setting a7), and 
then executing the syscall.

%% file: experiment.tex
\section{Experimentation on operational application}
\label{ref:experiment}

In order to experiment the feasibility of JOP attacks in the RISC-V architecture,
we designed an application emulating the behaviour of a sensor network
application running in an Industrial Control System (ICS).

The experiment prototype aims at monitoring the temperature of a real-time
critical system. In order to do so a pair of connected thermometers send
periodically encrypted payloads to the ICS. All sensors share the same
encryption key -- an existing architecture in an industrial system. The aim
of the attacker in our scenario is to steal the encryption key. With this key,
the attacker can perform malicious actions such as forging payloads to attack
the ICS.

We developped the target the application in C code and compiled it for RISC-V 
32bits, using gcc, disabling both Position-Independant Execution (PIE) and stack
protection (canaries), and with an optimization level of 1 for both application
and libraries. The target application was executed and validated on a RISC-V 
CV32A6 softcore design\footnote{https://github.com/openhwgroup/cva6}~\cite{zaruba2019cost} 
deployed on a Genesys2 FPGA running Linux.

In the following subsections, we explain how we were able to steal the key
within the application in the ICS using exclusively JOP.

\subsection{Performing the Attack}

\subsubsection{Attack Model}

The attack we realized in our experiment relies on a memory vulnerability,
for instance a buffer overflow, allowing to highjack the execution flow of
the target application. We did not activate memory randomization nor stack
protections techniques such as StackGuard\cite{261393} on the target 
application, since the former can be bypassed by different techniques~\cite{Shacham05}, 
and the latter is ineffective against JOP attacks.

We also made the hypothesis that, during the reconnaissance phase of the attack,
the attacker is able to access an exact twin of the target application (either
by rebuilding it with the same options and environment, or by acquiring an
device running said application), and run it with a debugger. Thus, potential
anti-debug, anti-reverse or anti-tampering measures are not considered within
the scope of our experiment.

\subsubsection{Attack overview}

The crafting of the attack consists in five steps divided between the two first
stages of the cyber kill chain: reconnaissance and weaponization.

The three steps involved in the reconnaissance stage are:
\begin{itemize}
  \item identification of a memory vulnerability;
  \item identification of gadget codes available in the application;
  \item definition of the attack aim.
\end{itemize}

The two steps involved in the weaponization stage are:
\begin{itemize}
  \item crafting of the JOP chain;
  \item crafting of the malicious payload.
\end{itemize}

As a first step the attacker must identify a memory vulnerability allowing to
hijack the execution flow toward the gadget initializer. This step is not covered
in detail in this document, but many techniques and tools exists in order to
identify such vulnerabilities~\cite{Younan04codeinjection}.

In our experiment, we inserted a vulnerability with the target application,
allowing a buffer overflow in the heap space. In order the vulnerability to
be exploitable for JOP attack, the buffer was collocated with a structure
containing functions pointers. The buffer overflow allows the overwriting of
said pointers, thus triggering an execution flow hijacking when they will be
called by the application -- so not necessarily just after the buffer writing.
Standard canaries techniques will not be able to block such exploit, since 
they only protect the stack.

\subsubsection{Identifying gadgets}

The identification of available gadgets in the application is the most important
stage in the attack, as it
decides which assets can be targeted by the attack. Too few, or not diverse
enough gadgets will restrict the range of available targets. While in our
experiment the gadgets were deliberately inserted in the application, in a
more realistic attack crafting the attacker would need to find them by
himself, which would indeed add a level of difficulty, although history of
gadget chain attacks suggests that tools for automated gadget detection may
emerge in a close future.

\begin{figure}
\includegraphics[scale=0.65]{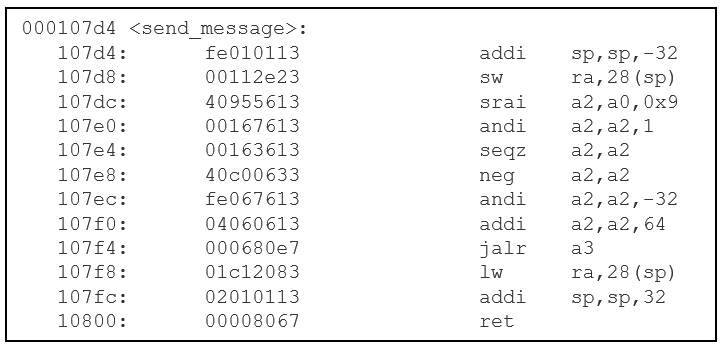}
\caption{Identifying gadgets}\label{fig:identify}
\end{figure}

In order to manually identify or confirm the presence of gadgets in the
target application, we used simple tools for machine to assembly language
translation such as objdump from the GCC suite. Since it is not necessary
for gadget identification to have a deep understanding of the application
behaviour, the use of tools that enable observation of the application at
runtime (such as debugger) is not mandatory at this stage. Figure~\ref{fig:identify}
illustrates the result of objdump on the target application binary.
The first line displays address and name (if the binary is not stripped)
of the disassembled function. Following lines consists in 3 columns:
\begin{itemize}
  \item instruction address;
  \item instructions (in hexadecimal format);
  \item translation of the instruction opcode and arguments in assembly language.
\end{itemize}

The attacker will focus on addresses, opcodes and arguments. The latter two
allow to identify code snippets of interest, that is the functional part
of the gadget. It must be followed by a chaining mechanism allowing to branch
to the gadget dispatcher address, through minimum register manipulation. In our
reserved registers approach where said gadget dispatcher is always stored into
the same register (a3), it means a JALR operation with said register in argument.

The addresses will be used in the JOP chain crafting step to build the dispatcher
table.

\subsubsection{Definition of attack objective}

The objective of the attack is to be decided from (1) system assets accessible
by the application and (2) available functional gadgets identified. The first
part requires acquiring a deep understanding of the application with techniques
such as runtime observation (e.g. through debugging, resource use, etc.) or
static analysis of the binary. Investigation on proprietary or domain-specific
protocols (quite prevalents in the industry) are also likely to be needed. The
second part typically requires domain-specific expertize (e.g. avionics,
railway, energy...), and knowledge of the technical environment (e.g. hardware,
middleware, connected devices...), especially in case of sabotage intent.

In our experiment, our objective is a AES256 secret key used for group communication
in the target sensor network system. Knowing this key allows a third party to
(1) read any message within the network and (2) forge fake messages. The latter
option makes sabotage action in the target ICS possible. Leaking said secret
key requires (a) arithmetic and logic gadgets, (b) memory access gadgets and
(c) system calls gadgets. These gadgets need to be consistent in the registers
use and free of side-effect on reserved registers. This hypothesis is
very conservative, and may be relaxed if register switching was possible.

\subsubsection{Design of the JOP chain}

Once the functional gadgets are identified and the objective of the attack is
defined, the actual gadget chain can be crafted. We proceeded in reverse
order to the attack: from the last operation of the attack, we inferred
the needed context in terms of registers and memory setting, and then choose
the gadgets allowing to set up this context.

The final operation of the attack will be a syscall (ecall instruction in
RISC-V) of type write, taking three arguments as parameters: register a0
containing the target file descriptor, register a1 containing the source
buffer address (the data to write), and register a2 containing the size of
the data to write.

In order to set up such a syscall, we had two main challenges: (1) find the
syscall address and (2) set up its parameters.

\begin{figure}
\includegraphics[scale=0.5]{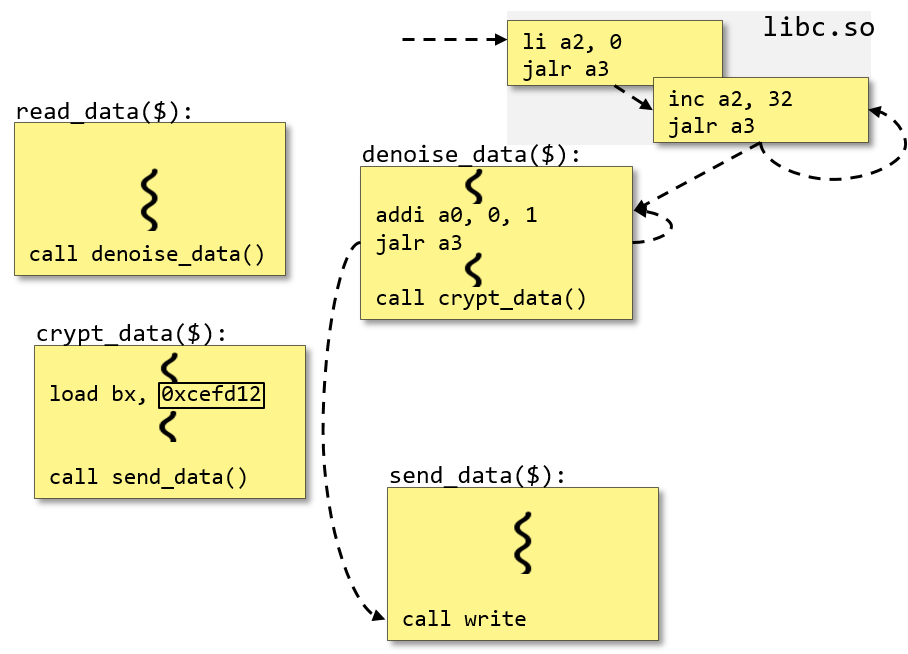}
\caption{JOP chain}\label{fig:jopchain}
\end{figure}

In our experiment we simply printed the key on the monitor, so a0 had to be
set to the stdout or stderr value. More realistic and useful attacks may target
an already-open file or socket. In order to achieve our goal, a gadget setting
a0 to 1 (the stdout file handler identifier) was used. We used a gadget available
from the reading operation in order to set a1 to the address of the secret key.
Finally, specifying the size of the writing was the most tricky part, since no
gadget was setting the literal value 256 into a2 -- the literal was actually not
present in the application. Thus, in order to implement the setting of a2, we
used two gadgets: one for the initialition of a2 to 0 and a second one to
increment of a2 to a literal value which was a divider of 256 (in our case, 32).
By calling the latter gadget multiple times (8 times), the final value of a2 was
256. These gadgets chaining is illustrated in Figure~\ref{fig:jopchain}.

Regarding the syscall address, we could find it by computing offline its offset
with an actually used function in the libc. This last snippet of code is
particular, as it does not need to branch back to the gadget dispatcher, since
the attack is already performed.

\subsubsection{Initialization of the attack}

Once the chain has been designed, the attack can be encoded in the dispatch table.
We use a sequential table, which address all the gadgets, including repetitions of
the same gadgets like in the case of incrementing of a2. The initializer gadget allows
setting a3 and a4, which contains respectively the address of the gadget dispatcher
and the address of the dispatch table. Finally, both the dispatch table and the initial
diversion to the gadget initializer will be embedded into a payload to be sent
through the attack vector. In our experiment, we targeted a non-protected buffer
contained within a structure in the in the heap, followed with a function pointer.

\subsection{Results and Limitations}

By using the techniques presented in this section, we were able
to steal the key embedded into the application's memory. Nevertheless we found
a limitation in our current approach: due to the flushing of the arguments
registers by system calls, we were not able to make more than one system call.
While it was enough for our attack objective, a more ambitious objective may
need to chain more than one function call -- for instance to read a file, and
then modify its content. This could be addressed by finding a way to
systematically store arguments before performing system calls, and restore them
after.

\subsection{Next Steps}

In our approach, we had a very conservative view on the surface attack, exploring
only manipulations of the arguments registers. Exploiting more registers -- such as
temporary registers -- may allow to find more available gadgets. Exploring commonly
used extensions may also be a new source of vulnerabilities. Using compressed
instructions would also allow shifting instructions and thus create new available
gadgets. Finally, finding a mechanism allowing to store and restore registers would also
increase the number of available gadgets, as some side effects can be overcome.

Our works proved that gadgets allowing JOP attacks can exist in RISC-V
architecture. One important step now is to evaluate how prevalent these gadgets
are in real-life applications, particularly in embedded systems. Furthermore, considering the
complexity of building gadget chains, automatizing the process will be necessary.

%% file: related.tex
\section{Related works}

\subsection{Experimentations on CRA attacks for RISC-V}

Brizendine et al.~\cite{joprocket21} proposed and implemented a method allowing 
building JOP gadgets chains for the x86 architecture. Similar effort could -- and most 
certainly will -- be done for RISC-V architectures.

Gu et al.~\cite{gu2020riscvrop} identified a specific pattern of instructions 
allowing linking functional gadgets in RISC-V architectures, introducing the 
concept of ``self-modifying gadget chain'' to save and restore registers value in 
memory. They also demonstrated the Turing-completeness of their solution. Adapting 
self-modifying gadget chain to JOP is indeed a promising solution to increase our 
capacity to build effective gadget chains. Jaloyan et al~\cite{Jaloyan20} reached 
the same result by abusing compressed instructions (\emph{overlapping}).

\subsection{Defenses from CRA for RISC-V}

Austin et al.~\cite{morpheus21} published the MORPHEUS II solution to RISC-V. This 
hardware-based solution aims at defeating memory probes trying to bypass address 
randomization by providing a reactive, fine-grain, continuous randomization of 
virtual addresses, as well as encryption of pointers and caches. This solution, 
while having a low overhead in terms of energy consumption and area, is quite 
intrusive in the hardware and may require efforts for certification in critical 
applications. While authors make no claim about stopping JOP attacks, probe-resistant 
ASLR may be difficult to bypass for an attacker.

Palmiero et al.~\cite{Palmiero18} proposed a hardware-based adaptation of Dynamic 
Flow Information Tracking (DIFT) for RISC-V, with the ability to detect most function 
pointers overwriting, whether direct or indirect, and in any memory segment, thus 
allowing blocking the attack at its initialization stage. Although this approach seems 
indeed powerful, it implies modification of RISC-V instructions behaviour (in extensions 
RV32I and RV32M), as well as in the memory layout (by adding a bit every 8 bits of 
memory). Such modifications, in addition to drifting away from RISC-V ABI, are likely 
to make certification difficult, a serious drawback in critical industrial systems. 
De et al.~\cite{heapsafe22} implemented a chip compliant to RISC-V, including a Rocket 
Custom Coprocessor (RoCC) which extends the RISC-V ISA with new instructions allowing 
safe operation on the heap. The authors ensure heap size integrity and prevent 
use-after-free attacks, at the cost of an increase of 50\% of average execution time 
on authors benchmarks~-- which is half than same solutions implemented at software 
level.

%% file: conclusion.tex
\section{conclusion}

Anticipating security vulnerabilities for RISC-V systems 
in order to identify and prevent possible attacks is an important challenge.
In this article, we demonstrated the feasibility of JOP attacks on RISC-V by 
exploiting vulnerable patterns in an application. Since we were conservative 
in our selection and usage of registers, the exploited patterns are likely to be 
a subset of actually vulnerable patterns. While we conducted a simple 
attack that performs only a single system call, extending the scope of exploited 
registers may allow  bypassing this limitation in the future. Register 
swapping or register saving techniques are also likely to be useful for that matter.
These techniques would indeed increase the complexity of crafting a 
consistent gadget chain. Thus, automated gadget detection and assembling techniques 
for RISC-V will be required, similarly to JOP Rocket tool~\cite{joprocket21}. With these tools 
and techniques~--- all within the domain of reasonable in short-term~--- an 
attacker would be able to perform arbitrarily complex and possibly stealthy 
attacks, as it would be possible to reset registers to a legit value after 
the exploit was performed.

In order to face these threats, security researchers need to devise solutions to 
address JOP attacks. Analyzing attack patterns, as we did in this experiment,
was a first, mandatory, step. Providing efficient, innocuous and low-overhead 
solutions suitable for critical embedded and real-time systems will be the next 
step.